\documentclass[11pt,a4paper]{article}

\usepackage{graphicx}
\usepackage{hyperref}

\begin{document}

	\title{AutoPass: An Automatic Password Generator }
	
\author{Fatma Al Maqbali and Chris J Mitchell\\
	Information Security Group, Royal Holloway, University of
	London\\
		\href{mailto:fatmaa.soh@cas.edu.om}{\nolinkurl{fatmaa.soh@cas.edu.om}},
	\href{mailto:me@chrismitchell.net}{\nolinkurl{me@chrismitchell.net}}
}
\date{15th July 2016}

\date{\today}

	\maketitle 	
\begin{abstract}
	Text password has long been the dominant user authentication technique and is used by large numbers of Internet services. If they follow recommended practice, users are faced with the almost insuperable problem of generating and managing a large number of site-unique and strong (i.e.\ non-guessable) passwords. One way of addressing this problem is through the use of a password generator, i.e.\ a client-side scheme which generates (and regenerates) site-specific strong passwords on demand, with the minimum of user input. This paper provides a detailed specification and analysis of AutoPass, a password generator scheme previously outlined as part of a general analysis of such schemes. AutoPass has been designed to address issues identified in previously proposed password generators, and incorporates novel techniques to address these issues. Unlike almost all previously proposed schemes, AutoPass enables the generation of passwords that meet important real-world requirements, including forced password changes, use of pre-specified passwords, and generation of passwords meeting site-specific requirements.
\end{abstract}

\section{Introduction}

Despite its widely-discussed shortcomings, text password authentication is widely used to authenticate users to online services. Many attempts have been made to replace simple password authentication, e.g.\ using biometrics, tokens and multi-factor authentication \cite{HerleyO12}. However, single-factor password-based authentication remains very widely used. Moreover, in recent years the number of widely-used password-protected services has grown significantly, in turn increasing the number of passwords users are expected to remember.
There are a range of issues associated with the use of text passwords \cite{FlorencioHO14}. These issues can be categorized as either user-related or online service-related.
User-related issues include:
\begin{itemize}
\item  users are likely to be overwhelmed by the large number of passwords needed for Internet services, which can lead to use of the same password for multiple accounts;
\item users will often choose guessable passwords, e.g.\ date of birth, name of pet, or anniversary date;
\item users will often make minimal modifications to an existing password, e.g.\ by including a serial number, when forced to make a change.
\end {itemize}

Online service-related issues, which can make it almost impossible for users to remember all their passwords, include:
\begin{itemize}
 \item many sites enforce a complex password policy, e.g.\ requiring passwords to contain a minimum number of characters, or include or exclude specific characters;
 
 \item some services force users to change their password regularly, e.g.\ every 90 days. 
\end{itemize}

To address these issues, a number of password generator schemes have been proposed \cite{cite6,cite4,cite2,cite5,Wolf06,cite7}, which generate strong (i.e.\ difficult to guess) and random-looking passwords and regenerate them whenever necessary. In a previous paper \cite{MaqbaliM16}, we evaluated existing password generator schemes and discussed their strengths and weaknesses; this analysis enabled us to outline a new scheme which we called \textit{AutoPass}. AutoPass is an on-demand password generator which generates site-specific passwords for online services. It combines features from existing password generators with novel techniques designed to address identified shortcomings in the existing schemes. In this paper we provide the first detailed specification of AutoPass, and also give a detailed analysis of its properties.

The remainder of the paper is structured as follows.  Section 2 defines what we mean by password generators, and gives a general model for such schemes (based closely on \cite{MaqbaliM16}).  Section 3 then gives a high-level description of AutoPass, building on the outline previously provided.  This is followed in section 4 by a detailed specification of the operation of AutoPass.  Section 5 provides an analysis of the properties of AutoPass, and in particular highlights how it addresses known shortcomings of such schemes. Finally, the paper concludes in section 6.

\section{ A General Model} \label{model}
\subsection{Definition}
This paper is concerned with \emph{password generators}, i.e.\
schemes designed to simplify password management for end users
by generating site-specific passwords on demand from a small
set of readily-memorable inputs.  Note that the term has also
been used to describe schemes for generating random or
pseudorandom passwords which the user is then expected to
remember; however, we use the term to describe a system
intended to be used whenever a user logs in and that can
generate the necessary passwords on demand and in a repeatable
way. A variety of such schemes have been proposed in recent years.

In this section we present a general model for such
schemes, which we use later in this paper as the basis for describing our novel scheme, AutoPass \cite{MaqbaliM16}. We observe that the general class
of such schemes has been briefly considered previously by
McCarney \cite{cite3} under the name \emph{generative password managers}.

\subsection{A Model}
 A password generator has the following components.

\begin{itemize}
\item A set of \emph{input values} is used to determine the
	password for a particular site.  Some values must be
	site-specific so that the generated password is
	site-specific.  The values could be stored (locally or
	online), based on characteristics of the authenticating
	site, or user-entered when required. Systems can, and
	often do, combine these types of input.
\item A \emph{password generation function} combines
	the input values to generate an
	appropriate password.  This function could operate in a
	range of ways depending on the requirements of the web
	site performing the authentication. For
	example, one web site might forbid the inclusion of
	non-alphanumeric characters in a password, whereas
	another might insist that a password contains
	at least one such character.  To be broadly
	applicable, a password generation function must
	therefore be customisable.
\item A \emph{password output method} enables the generated
	password to be transferred to the authenticating site.
	This could, for example, involve displaying the
	generated password to the user, who must then type (or
	copy and paste) it into the appropriate place.
\end{itemize}

All this functionality needs to be implemented on the user
platform. There are various possibilities for such an
implementation, including as a stand-alone application or as a
browser plug-in.

\subsection{Examples}

Before proceeding we briefly outline some existing proposals
for password generation schemes conforming to the above model.
The schemes are presented in chronological order of
publication. 	
\begin{itemize}
\item The \emph{Site-Specific Passwords (SSP)} scheme
	proposed by Karp \cite{cite4} in 2002/03 is one of the
	earliest proposed schemes of this general type.  SSP
	generates a site-specific password by combining a
	long-term user master password and an easy-to-remember
	name for the web site, as chosen by the user.
\item \emph{PwdHash}, due to Ross et al.\ \cite{cite5},
	generates a site-specific password by combining a
	long-term user master password, data associated with
	the web site, and (optionally) a second global password
	stored on the platform.
\item The 2005 \emph{Password Multiplier} scheme of
	Halderman, Waters and Felten, \cite{cite6}, computes a
	site-specific password as a function of a long-term
	user master password, the web site name, and the user
	name for the web site concerned.
\item Wolf and Schneider's 2006 \emph{PasswordSitter}
	\cite{Wolf06} generates a site-specific password
	as a function of a long-term user master password, the
	user identity, the application/service name, and some
	configurable parameters.
\item \emph{Passpet}, due to Yee and Sitaker \cite{cite7}
	and also published in 2006, takes a very similar
	approach to SSP, i.e.\ the site-specific password is a
	function of a long-term user master password and a
	user-chosen name for the web site known as a
	\emph{petname}.  Each petname has an associated icon,
	which is automatically displayed to the user and is
	intended to reduce the risk of phishing attacks.
\item \emph{ObPwd}, due to Mannan et al.\
	\cite{Biddle11,cite2,Mannan08,cite1}, first surfaced in
	2008.  It takes a somewhat different approach by
	generating a site-specific password as a function of a
	user-selected (site-specific) object (e.g.\ a file),
	together with a number of optional parameters,
	including a long-term user password (referred to as a	\emph{salt}), and the web site URL.
\item Finally, \emph{PALPAS}
	\cite{horsch15} generates passwords complying
	with site-specific requirements using server-provided
	password policy data, a stored secret master password
	(the \emph{seed}), and a user-specific secret
	value (the \emph{salt}) that is synchronised across all
	the user devices using the server.
\end{itemize}

There are also widely available applications conforming to the general model, examples of which we briefly mention below.  The following schemes are available as browser extensions.

\begin{itemize}
	\item \textit{RndPhrase}\footnote{https://rndphrase.appspot.com/} is a Firefox add-on and web-based password generator.  It generates site-specific passwords as a function of a predefined salt (unique per user), the host name, and a user-entered password.  The user only needs to remember the password.
	
	\item  \textit{PwdHash port}\footnote{\url{https://addons.opera.com/en-gb/extensions/details/pwdhash-port/}} is an Opera add-on based on PwdHash.
	
\end{itemize}
 Android Phone App:
The following two apps are available for Android mobile phones.
\begin{itemize}
	\item \textit{Advanced password generator}\footnote{\url{https://goo.gl/MF0z1D}} generates passwords which can be copied and shared.  The form of a generated password, e.g.\ its length and character(s) to include/exclude, can be configured.

	\item \textit{Password generator}\footnote{\url{https://goo.gl/SNVtJY}} generates passwords as a function of a configurable set of parameters including a user salt.  The system indicates its estimate of the ‘strength’ of the generated password.
\end{itemize}

\subsection{Registration and Configuration}
\label{registration} \label{configuration}

In this paper we are interested in schemes whose operation is
completely transparent to the website which is authenticating
the user. As a result, the `normal' website registration
procedure, in which the user selects a password and sends it to
the site, is assumed to be used.  This, in turn, typically
means that password generation needs to take place
\emph{before} user registration with a site, or at least that
introduction of the password generator requires the user to
modify their account password. This requirement causes problems with all the previously proposed password generator schemes, as we discuss in section 2.5 below.

There is a potential need for a password generator to store
configuration data. Such data can be divided into two main
types:
\begin{itemize}
\item \emph{user-specific configuration data}, i.e.\ values unique
	to the user and which are used to help generate all
	passwords for that user, e.g.\  a master password; and
\item \emph{site-specific configuration data}, i.e.\ values
	used to help generate passwords for a specific website,
	which are typically the same for all users, e.g.\ a password policy.
\end{itemize}
Not all schemes use configuration data, although producing a
workable system without at least some user-specific configuration data
seems challenging. However, the use of configuration data is
clearly a major barrier to portability. That is, for a user
employing multiple platforms, the configuration data must be
kept synchronised across all these platforms, a non-trivial
task --- exactly the issue addressed in a recent paper by
Horsch, H\"{u}lsing and Buchmann \cite{horsch16}.
\subsection{Issues with existing schemes}
Before describing the detailed operation of AutoPass, we observe certain fundamental problems that affect all (or almost all) previously proposed password generators. These issues motivate the design of AutoPass, which incorporates novel features designed to overcome these problems.
\begin{description}
\item[Setting and updating passwords] As noted above, if a user is already using the password generator when newly registering with a website, there is clearly no problem --- the user can simply register whatever value the system
generates. However, if the user has selected and registered passwords
with a range of websites before starting use of the password generator,
then all these passwords will need to be changed to whatever the
password generator outputs. — This could be highly inconvenient if a
user has established relationships with many sites, and could present
a formidable barrier to adoption of the system.

Somewhat analogous problems arise if a user decides to change a website
password, e.g.\ because the site enforces periodic password changes.
The only possibility for the user will be to change one of the inputs
used to generate the password, e.g.\ the object (if a digital object is
used as an input) or a user site name. Password change could even be
impossible if the user does not choose any of the inputs
used to generate a password.
\item[Using multiple platforms] If a user employs multiple platforms, e.g.\ a desktop and a smart phone, then problems will arise if any locally-stored configuration data is used.
\item [Password policy issues] A further general problem relates to the need to automatically generates passwords in a site-specific form, a problem not satisfactorily addressed by any of the previously proposed schemes except PALPAS\@. Some existing  schemes have the option for the user to customise a generated password, but the user has to identify the requirements for the website manually and configure the options accordingly. Automatically generating a password tailored to meet a website’s specific requirements has been explored extensively by Horsch and his co-authors \cite{Mannan08,cite1}. 
\end{description}

\section{AutoPass: Instantiating the Model}

\subsection{ AutoPass Components}

AutoPass has two main components: the AutoPass server and the AutoPass client software. The AutoPass server is used to store relatively non-sensitive user data, e.g.\ the user name and website-specific password policies (i.e.\ specifications of the types of password a particular site will accept). The AutoPass client software provides a user interface, and automatically generates site-specific user passwords as a combination of the specified set of inputs.  Some of the inputs are stored locally and some are stored in the AutoPass server, with which the client software interacts as necessary. Where possible, the generated password is automatically inserted into login forms.

\begin{figure}[h!]
	\includegraphics[width=\linewidth]{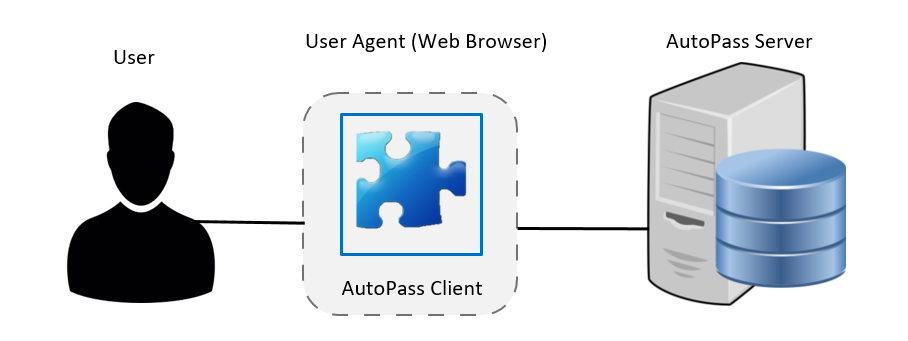}
	\caption{ AutoPass System Architecture}
	\label{ AutoPass System Architecture}
\end{figure}

Figure \ref{ AutoPass System Architecture} depicts the AutoPass architecture, showing the main components of the scheme.

\subsection{Overview}
We next provide a high level description of AutoPass, based on the general model provided in the previous section. We first describe the three main components of the scheme, i.e.\ the input values, the password generation function, and the output method, together with an initial approach to implementation.

\begin{itemize}
\item \textbf{Input values}.  We propose the use of a range of types of input, as listed below, incorporating those used in previous schemes.

\begin{itemize}
	\item A \textbf{master password}, i.e.\ a long-term strong password selected by the user or generated by the system. The master password is stored in encrypted form on the AutoPass server as part of the user-specific configuration data for that user. Since it does not need to be remembered by the user, this could, for example, be a 128-bit random value. The precise choice is implementation-dependent.  The user might wish to make a written record of this value when it is initially chosen, and store it securely for backup/recovery purposes.
	
	\item The \textbf{site name} is the URL of the site for which the software is generating a password.  To overcome the issue of changes to URL sub-domains, AutoPass only uses the first part of the URL (i.e.\ up to the first / character ).
	
	\item A \textbf{password policy} specifies the set of site-specific requirements for a password (e.g.\ including a length constraint and/or setting minimum numbers of certain classes of character).  Many websites enforce highly specific policies, reflecting somewhat ad hoc decisions made by system designers. The policy is specified using the Password Requirements Markup Language (PRML) \cite{horsch16}.
	
	\item A \textbf{digital object} is a text fragment, picture, or audio sample, typically in the form of a digital file.  This is an optional input that potentially adds significant entropy to the password generation process, e.g.\ for use when generating passwords protecting high-value resources.	
\end{itemize}
 
 \item In the \textbf{password generation} stage, the input values are combined to produce the desired site-specific password. This process occurs in two stages, as follows.
 \begin{itemize}
 \item The first stage involves combining the input values, including the master password and the URL, to produce a bit string.  Following Kelsey et al.\ \cite{Kelsey}, this computation involves a two-level hash computation, as follows.
 \begin{enumerate}

 \item 	The site-independent input (namely the master password) is submitted to a cryptographic hash-function, e.g.\ SHA-256, \cite{ISOSHA}. The hash-function is iterated a significant number of times, e.g.\ 1000, where the number is chosen to be as large as possible without making the client software too unresponsive.  This number can be  user-dependent, as long as the value is held in the server to enable it to be synchronised across all the platforms for a particular user.  The output, e.g.\ a 256-bit string, is then cached in the client software.  Since this value is independent of the website, it can be computed once when the client software is started up and cached locally while the client software is active.
\item The 256-bit string is concatenated with the website-specific inputs (the site name and the optional digital object) and hashed once more using the same hash-function to yield a site-specific bit string.
\end{enumerate}
 The use of a two-level process is designed to give some protection against brute force attacks by slowing them down.  Suppose an opponent knows a site-specific password and wishes to use this to learn the user's master password by searching through all possibilities.  Of course, if the master password is a randomly chosen 128-bit string then there is no danger of such a search succeeding, but some users may not select their master passwords to possess high entropy.  In such a case, it might be possible to work through all possibilities for the master password until the correct value is found.  However, use of a multiply-iterated hash-function means that such a brute force search will involve significantly more computational effort than it otherwise would; at the same time, since the iterated part is only computed once per session, the additional load on the genuine client will be manageable.
 \item The second stage (encoding) involves taking the bit string output from the first stage and constructing from it a password of the desired form, using the PRML policy specification to ensure the password meets the website-specific requirements.  Other possible inputs to this encoding step include the password offset.  How encoding  operates is discussed in detail in section 3.4 below.
 \end{itemize}
  \item  \textbf{Password output and use} is achieved by automatically copying the generated password to the targeted password field. AutoPass uses secure filling techniques to prevent sweeping attacks \cite{pwdmgr2014}.

\end{itemize}
  We propose to implement AutoPass (at least initially) as a browser add-on.  This will enable us to automate key tasks, including fetching the website URL and inserting passwords into login forms.  In the future, for use with platforms not permitting such add-ons, we plan to examine both stand-alone applications and web-based functionality.

\subsection{ Stored Data}
As we have described, AutoPass needs access to a variety of configuration data to enable it to operate automatically, and this configuration data clearly needs to be stored somewhere. There are two possible locations for data storage, namely the AutoPass server and the AutoPass client, and AutoPass uses both. The configuration data stored at the server is held long-term, i.e.\ for the lifetime of the user account at the server; the data held on the client may be held either short-term, typically for the life of a session, or long-term, i.e.\ while the software remains installed on the client platform.  We summarise below the various types of stored data.

\subsubsection{Server-stored data}
The following user-specific configuration data is held at the AutoPass server:
\begin{itemize}
	\item	the user account name;
	\item	an email address for the user;
	\item	the (encrypted) master password;
    \item  	a hash of the master password;
	\item   a (salted) hash of the session password
    \item	for each website for which a password has been generated for this user:
	
	\begin{itemize}
	\item	the (first part) of the URL of the website;
	\item	the types of input used to generate the password for this site;
	\item   the password offset for this site (see section 3.5)
\end{itemize}

\end{itemize}

The following site-specific configuration data is held at the AutoPass server:
\begin{itemize}
	\item	the (first part) of the URL of the website;
	\item	the password policy of the site, encoded in PRML (see section 3.4);
\end{itemize}

Note that the site-specific configuration data could be maintained by a server separate from that used to store the user-specific configuration data.  Indeed, since this data is completely non-confidential, it could be provided by a service independent of AutoPass, e.g.\ the Password Requirements Description Distribution Service (PRDDS)\cite{horsch16}, which provides  an online interface to meet requests for PRML-based Password Requirements Descriptors (PRDs) for websites identified by their URL.

\subsubsection{Client-stored data}
The following data is held long-term by the AutoPass client:
\begin{itemize}
	\item	cached copies of password policies for recently visited websites.
\end{itemize}

The following data is held short-term by the AutoPass client:
\begin{itemize}
	\item	the session password;
	\item	a multiply-iterated hash of the master password (see section 3.2).
\end{itemize}

\subsection{ PRML}

The Password Requirements Markup Language (PRML) \cite{horsch16} is an XML-based syntax that can be used to specify site-specific password requirements, including minimum and maximum lengths, the permissible character set, and minimum required number(s) of specific sub-classes of characters. It has been designed to address the diversity of password requirements arising in practice, and enable password generators to automatically generate site-specific passwords that match online service password requirements.

The PRML specification for a website provides one of the two inputs for the second stage of password generation described in section 3.2, the other being the bit string output from the first stage.  The second stage of password generation operates as follows.

\begin{enumerate}
\item The size $C$ of the password character set is derived from the PRML specification; we suppose that a mapping is chosen from the set of integers $ \lbrace {0,1,\ldots,C-1}  \rbrace  $ to the characters in the password character set.

\item The length $L$ of the password is chosen to be the minimum of 16 and the minimum length prescribed by the PRML policy.

\item 	The input bit string is converted to a positive integer (by regarding the string as the binary representation of a number), and this number is converted to its $ C $-ary representation $d_{t}d_{t-1}\ldots d_{0}$, for some $  t $, where $ 0\le d_{i}\le C-1$  for every $ i $  $ (0\le i\le t) $.

\item The final  $  L $ digits of the above sequence of numbers, i.e.\  $d_{L-1}d_{L-2}\ldots d_0$, are then converted to characters using the mapping established in step 1.

\item The password is tested to verify that it satisfies the other constraints in the PRML specification.  If not, then the input bit string is rehashed and the process is recommenced; otherwise the process is complete.
\end{enumerate}

The above procedure operates on the assumption that the length of the input bit string is significantly greater than $ \lceil  L \log_{2} C \rceil$.  Since the likely value of $ L $ is 16, and a typical value for $ C $ is at most 64, this means that $ \lceil  L \log_{2} C \rceil$ is likely to be less than 100, i.e.\ much less than the length of the output of a modern hash function such as SHA-256  (which gives a 256-bit output).  It is also based on the assumption that a ‘random’ password with characters from the specified password set has a reasonable chance of satisfying the PRML requirements.  If this latter assumption is not true, then a more elaborate second stage algorithm could be devised.

\subsection{Password Offsets}

As noted in section 2.5, one major issue with existing password generation schemes is that they do not provide a facility for a user to choose a password (e.g.\ to allow continuing use of a password established prior to use of AutoPass), or to change a password without changing the set of inputs.  We propose the use of \textit{password offsets} to support these requirements.  A password offset works in the following way.
\begin{enumerate}

\item	A password $d_{L-1}d_{L-2}\ldots d_{0}$ is first generated in the normal two-stage way (as described in section 3.4), and suppose $  D $ is the positive integer which has $d_{L-1}d_{L-2}\ldots d_{0}$ as its $ C $-ary representation.
\item 	Let the user-chosen password (of length $ M $, say) be encoded as a sequence of $ M $ digits $ e_{M-1}e_{M-2}\ldots e_{0} $, where $  0\le e_{i}\le C-1 $ for every $ i$ $  (0\le i\le M-1) $, and suppose $ E $ is the positive integer which has $ e_{M-1}e_{M-2}\ldots e_{0} $ as its $ C $-ary representation.
\item 	The password offset is simply $ E-D $.
\end{enumerate}
When generating a password, if a password offset exists then the password can simply be generated in the normal way and the offset ‘added’, and the result will be the $ C $-ary encoding of the desired password.  A similar approach can be used to force a password change, where a new password can be generated at random (in accordance with the PRML specification) and the password offset can be set to the difference between the new password and the value generated using the standard procedure.

\section{Details of Operation}

\subsection{General}

We next provide details of the operation of AutoPass, including: application installation and setup, operational sessions, initial use with a website, and subsequent use with a website. Note that, to simplify the discussion, we assume that AutoPass is implemented as a browser add-on running on a Windows platform. Alternative implementation scenarios, e.g.\ as a stand-alone application on a phone or tablet, are likely to be very similar, but may vary in some minor details.

\subsection{ Application Installation}

We divide this discussion into two cases, i.e.\ where a user installs AutoPass for the first time (and creates an account on the AutoPass server), and where a user installs the software on an additional platform and already has an AutoPass account.

\subsubsection{First Installation and Account Creation}

We first describe the case where a user decides to start using AutoPass, and wishes to install and set up the client software. Once the AutoPass add-on is installed, the ensuing set-up procedure involves the following steps.\label{login}
 \begin{enumerate}
\item When the AutoPass add-on is activated for the first time, e.g.\ when the user clicks a toolbar button, it first asks the user whether he/she has an existing account.  In this case the user indicates that a new account is to be created.  Creating a new account  involves the client software contacting the AutoPass server, and various registration details need to be completed.  In particular, registration involves gathering the following pieces of information: \textit{session password}, \textit{user name}, and \textit{master password}.
	\begin{enumerate}
\item Session password: This is chosen and entered by the user, who must memorise it. It serves two main purposes: user authentication and  the derivation of a key used for encrypting the copy of the master password stored at the server. After entry of the session password, the client computes a salted hash of the value, which is sent to the AutoPass server as a means of authenticating the user in later sessions.

\item User name: The user must select a unique name. The AutoPass server checks that the name is not already in use, and if necessary requests the user to choose a different value. The user name is stored by the AutoPass server, and serves as an identifier for the user.

\item Master password: This value, which essentially functions as a cryptographic key, can be generated by the user or the AutoPass client software (perhaps at the choice of the user).  We assume for the purposes of this description that the master password is a 128-bit value, represented as a string of 32 hexadecimal characters.  If the client software generates it, it should be displayed to the user and the user should be advised to keep a copy somewhere secure so that system recovery is possible (see section 4.2.3 below). The session password is used to generate a cryptographic key, e.g.\ by hashing the concatenation of the session password and a fixed value. This key is then used to encrypt the master password using an appropriate symmetric encryption technique, e.g.\ AES \cite{ISOAES} in an authenticated encryption mode, prior to uploading it to the server.  The server retains this encrypted master password for downloading to a client whenever a user logs in.  The AutoPass client also generates a hash of the master password, which is sent to the AutoPass server and stored with the encrypted copy; – this hash value is only used for recovery purposes (see section 4.2.3).

\item Other information:  To complete registration, and to allow for recovery in the event of a user forgetting his or her user name or password, an email address (or addresses) should also be collected.  Other contact details could also be given, e.g.\ a mobile number.  The server holds this information as part of the user account information.
\end{enumerate}
			
\item After successful user account creation, the user is requested to log in using his or her newly established user name and session password (see section 4.3).	
\end{enumerate}
	
\subsubsection{Installing AutoPass on a newly Acquired PC}

Once an AutoPass account has been established (as described immediately above), the following steps are followed to set up AutoPass on a new machine.  We suppose that the client software has been installed on the platform.
	
		\begin{enumerate}
			\item As previously, when the AutoPass add-on is activated for the first time, it first asks the user whether he/she has an existing account. In this case the user indicates that he/she already has an account. The AutoPass client then asks the user for his or her user name and session password, and the process continues exactly as in a normal operational session (see section 4.3).	 
		\end{enumerate}

\subsubsection{Recovery}

The system needs to provide a recovery mechanism for the case where a user forgets their user name and/or password.  We suppose that the client software has a recovery function, which a user can invoke in the event of a forgotten user name or password.  We consider the operation of this recovery function for the two cases separately.

\begin{itemize}
	\item If a user forgets their user name, he or she can request a copy from the server by entering their registered email address.  The server checks that the email address is registered, and then sends the user name for this address to the user via email.
	
	\item If a user forgets their session password, then it cannot be recovered since neither the server nor the client retain a copy.  However, if the user has kept a copy of the master password, then system recovery is possible using the following procedure.  The user is prompted to enter his or her user name and the master password.  The user is also prompted to select a new session password.  The session password is used to generate a key which is used to encrypt the master password, exactly as described in section 4.2.1.  A hash of the newly selected session password, the user name, the encrypted master password, and a hash of the master password are all sent to the AutoPass server.  The AutoPass server authenticates the user by comparing the master password hash with its stored value, and if successful replaces the current encrypted master password and session password hash with the newly supplied values.  Finally, the server communicates the success of the recovery operation to the AutoPass client, which informs the user.
\end{itemize}
	
\subsection{Operational Sessions}

We next consider what occurs when the AutoPass software is activated, e.g.\ after the host platform has been rebooted. We suppose that the set-up process described in section 4.2 has already been performed.

\begin{enumerate}
\item	The user is prompted for his or her user name and session password.
\item	The user name is sent to the AutoPass server, which responds with the salt value for its stored copy of the hashed session password for the identified user.
\item	The AutoPass client software uses the salt value to hash the session password entered by the user, and the resulting hash-value is sent to the server.
\item	The AutoPass server checks that the session password hash is the same as its stored value, and by doing so authenticates the user.
\item	The AutoPass server sends back to the client the encrypted master password.  The AutoPass server also sends the following information for each site for which the user has created a password using AutoPass:
\begin{itemize}
	
\item the first part of the URL of the site (this is used as the site identifier);
\item	the password policy for the site (in PRML);
\item	the set of input types used to generate the password for this site (e.g.\ whether a digital object is used);
\item	the password offset for this site, if it exists;
\item	any other parameters used to control the generation of the password for this site.
	
\end{itemize}

\item	The AutoPass client decrypts the master password using a key derived from the session password, and multiply hashes the master password; the result is cached and the master password can then be deleted.
\item	Once activated, the AutoPass add-on will run continuously in the background,
examining each web page to see if it is a login page. It does
this by using various heuristics, including looking for the string \textit{input type=``password''}.
\item 	The add-on will then work as required, generating passwords automatically, until the session ends, e.g.\ when the browser is terminated.
\end{enumerate}

\subsection{ Initial Use with a New Website}

When AutoPass is used with a new website, the following procedure is executed.  We suppose that the AutoPass software is already executing, i.e.\ the procedure in 4.3 has been followed.

\begin{enumerate}
	\item	If the AutoPass add-on detects a login page, it cross-checks the first part of the site URL with the data downloaded from the AutoPass server to determine whether it is a site for which it has already generated a password.  In this case, we suppose that it has not previously been used to generate a password for this site.
	
	\item	The AutoPass add-on then communicates with the user by some means (e.g.\ via a pop-up) to indicate that it has detected a login page for a website for which a password has not previously been generated, and asks the user whether it would like AutoPass to manage generation of a new password for this site.
	
	\item	If the user declines, then the AutoPass add-on goes back to looking for login pages.  If the user accepts, AutoPass next asks what types of input the user would like to use to generate the password from amongst those listed in section 3.2.
	
	\item  The user selects the types of input to be used and, if the use of digital objects is selected, the user is also asked to select such an object.  The AutoPass client assembles the set of inputs, including the first part of the website URL and the multiply-hashed master password, to be used to generate the site password.  The AutoPass client also offers the user the option to select the password --- if the user requests this option then the user is prompted for the pre-chosen value.
		
	\item The password is generated using the procedure specified in sections 3.2 and 3.5 and automatically copied to the password field. If the user chose to select the password value, then the appropriate password offset is computed during password generation.
	
	\item The user preferences and the password offset (if appropriate) are sent to the AutoPass server for storage.
\end{enumerate}

\subsection{Subsequent Use with a Website (Everyday Use)}

The following steps are executed when AutoPass is in everyday use, i.e.\ after a website has already been set-up, as in section 4.4.  As in section 4.4, we suppose that the client AutoPass software is already active.

\begin{enumerate}
	\item 	If the AutoPass add-on detects a login page, it uses the first part of the site URL to check whether a password has previously been generated for this site --- in this case we suppose it has.
	\item 	The AutoPass add-on then assembles the set of inputs to be used to generate the password; if the user preferences for this site indicate that a digital object is to be used, the add-on prompts the user for the object.
	\item	The AutoPass add-on then generates the password, using the password offset if available, and automatically copies the generated value to the password field.	
\end{enumerate}

\subsection{Other Aspects}

\subsubsection{Client-server communications}

Whilst the data exchanged between AutoPass client and server is not necessarily highly confidential, some is privacy-sensitive and the integrity of all the data is crucial for correct operation.  We therefore propose that all data exchanged between client and server is protected using a server-authenticated TLS channel established at the beginning of a client session.  To make the authentication of server to client robust, the client is assumed to use certificate pinning for the server.

\subsubsection{Client caching}

The set of website data downloaded by the AutoPass server to the AutoPass client at the beginning of every session (see step 5 of section 4.3) is not likely to change very rapidly.  It therefore makes sense for the client to cache the most recently downloaded copy of this data, potentially improving system availability even if the AutoPass server is unavailable for a period.

\section{Evaluation}
We next consider the security properties of the scheme.
\subsection{Trust relationships}
Clearly, the AutoPass server must be trusted to some extent by the user, since if it sends incorrect data then passwords cannot be generated correctly.  It also learns which websites the user interacts with, and hence it must be trusted to respect user privacy.  On the other hand, it does not have the means to learn user passwords, since it only has access to an encrypted copy of the master password, which is used to generate all passwords, and thus it can be regarded as being ‘partially trusted’.

\subsection{Threat model}
The security and correct operation of AutoPass depends on a number of key assumptions, which we enumerate.
\begin{itemize}
\item 	The client device is assumed to be uncompromised, since passwords are generated in and used by this device. If, for example, the browser is compromised then clearly the generated passwords may be compromised.
\item 	The AutoPass client is assumed to be correct and without exploitable vulnerabilities.  As for the previous assumption, if a corrupted version of the client software is present on the client device, then user passwords may be compromised.
\item 	The AutoPass server provides correct information (see also section 5,1 above).
\end{itemize}

Given the above assumptions, AutoPass is designed to resist the following types of attack:
\begin{enumerate}
\item 	active attacks on the communications link between the AutoPass client and server, including masquerading as the server to the client or vice versa, e.g. as made possible by an untrustworthy wireless access point or by DNS poisoning;
\item 	attacks on password secrecy conducted by the AutoPass server;
\item 	attacks on password secrecy conducted by a valid site against user passwords for other sites;
\item 	attacks on password secrecy conducted by any party with temporary access to the AutoPass server database.
\end{enumerate}

\subsection{Security properties}
We conclude this discussion by considering whether the desired properties are realised by the AutoPass design.  We consider the four attacks in the previous section in turn.
\begin{enumerate}

\item 	Attacks of the first type are prevented by the assumption that all communications between the client and server are TLS-protected.  The server will be authenticated by a pinned certificate.  The client end of the link will not be explicitly authenticated to the server, but the presence of the correct user is verified by checking that the correct hash of the session password is sent over the link.

\item 	The server only has access to password metadata, a hash of the master password, and an encrypted copy of the master password.  If the master password is automatically generated by a process using sufficient randomness, use of a 128-bit value will prevent direct brute forcing guessing attacks.  However, the encryption of the master password is based on a key derived from the user-selected session password.  If the session password is poorly chosen, then it can be brute-forced, meaning that the server could gain access to the master password.  Thus it is vital for the user to choose a session password with high entropy.  This is a reasonable assumption since this is the only secret the user is required to memorise.
It is also worth noting that the use of password offsets means that if a password (and its offset) are compromised then all future passwords for that particular site can be determined if the offset is known.  That is, whilst offsets will not be divulged to any party, a dishonest AutoPass server that (by some means) learns a user’s password for a website will be able to determine all future passwords for that site.

\item 	The AutoPass system is completely transparent to authenticating websites.  If a website guesses that AutoPass is in use, it could use the password to try to perform a brute force search for the master password.  However, if the master password is chosen at random then such a search is infeasible.

\item 	If an unauthorised party has access to the AutoPass server database, then it will not have immediate access to any user passwords.  However, as discussed under 2 above, if the unauthorised party obtains the encrypted master password and the session password used to encrypt the master password is poorly chosen, then the attacker might be able to brute-force the session password and learn the master password.  This argues in favour of the AutoPass server providing additional encryption of the database, giving protection against compromise of stored user data.
\end{enumerate}

\section{Concluding Remarks}
In this paper we have described in detail the design of the AutoPass password generator.  Its use of a server, in particular to store password offsets, PRML specifications and user preferences, avoids the shortcomings present in all previously proposed password generators.  At the same time, this server is only partly trusted, and does not have the means to recover individual user passwords, as discussed in section 5.
Of course, whilst the system works in theory, it remains to verify that the system will work in practice.  A prototype implementation of AutoPass is being developed, and this will be used to conduct user trials, to verify that the desired high level of usability can be achieved.  We hope to report on these trials in a future paper.

\bibliographystyle{plain}
\bibliography{AutoPass}

\end{document}